\documentclass[onecolumn]{revtex4}

\usepackage{graphicx}
\usepackage{dcolumn}
\usepackage{bm}

\input epsf

\begin{document}

\title{\Large Study of Thermodynamics in Generalized Holographic and Ricci Dark Energy Models}

\author{\bf Samarpita Bhattacharya\footnote{samarpita$_{-}$sarbajna@yahoo.co.in}
and Ujjal Debnath\footnote{ujjaldebnath@yahoo.com ,
ujjal@iucaa.ernet.in}}

\affiliation{Department of Mathematics, Bengal Engineering and
Science University, Shibpur, Howrah-711 103, India.}

\date{\today}

\begin{abstract}
We have considered the flat FRW model of the universe which is
filled with the combination of dark matter and dark energy. Here
we have considered two types of dark energy models: (i) Generalized
holographic and (ii) generalized Ricci dark energies. The general
descriptions of first law and generalized second law (GSL) of thermodynamics have studied
on the apparent horizon, particle horizon and event horizon of the universe.
We have shown that the first law and GSL are always valid
on apparent horizon and first law can not be satisfied on the particle and
event horizons in Einstein's gravity. These results are always true for any
types of dark energy models i.e., these results do not depend on the dark energy
models in Einstein's gravity. But the GSL completely depends
on the choices of dark energy models in Einstein's gravity.
Here we have discussed the validity of GSL in Generalized holographic
and generalized Ricci dark energy models. On the particle horizon GSL may be satisfied
but on the event horizon the GSL can not be satisfied for both the dark energy models.
Also we have considered the Generalized holographic dark energy and generalized Ricci
dark energy as the original holographic dark energy, so in this situation we have
calculated the expression of the radius of the horizon $L$. On this horizon, we have
shown that the first law can not be satisfied. Finally, on the horizon of radius $L$,
we have found that the GSL can not be satisfied for both the dark energy models.\\
\end{abstract}


\maketitle

\section{\normalsize\bf{Introduction}}

Recent astrophysical data from distant Ia supernovae observations
[1, 2] show that the current Universe is not only expanding, but
also it is accelerating due to some kind of negative-pressure form
of matter known as dark energy [3, 4]. This mysterious fluid is believed to dominate over the
matter content of the Universe by 70 $\%$ and to have enough negative
pressure as to drive present day acceleration. However, the nature of dark energy
is still unknown, and people have proposed some candidates to
describe it. There are different candidates to play the role of the dark energy. The most traditional
candidate is a non-vanishing cosmological constant which can also be
though of as a perfect fluid satisfying the equation of state $p=-\rho$.
The next simple model proposed for dark energy is the quintessence [5]
which is a dynamical scalar field which slowly rolls down in a
at enough potential. There are different alternative theories for the dynamical dark energy scenario which have been
proposed by people to interpret the accelerating universe like K-essence [6], tachyonic field [7],
Chaplygin gas [8], phantom field [9], etc. Next we will discuss another two types of dark energy candidates i.e., holographic
and Ricci dark energies.\\

In quantum field theory a short distance (UV) cut-off is related
to a long distance (IR) cut-off $L$ due to the limit set by
formation of a black hole, which results in an upper bound on the
zero-point energy density [10]. It is to be noted that the size
$L$ should not exceed the mass of a black hole with the same size
[11], i.e. $L^3 \Lambda^4 \leq L M_p^{2}$ where $\Lambda$ is the
ultraviolet (UV) cut-off of the effective quantum field theory.
The largest IR cutoff $L$ is chosen by saturating the inequality,
so that the holographic dark energy density [12] may be defined by
$\rho_{\Lambda}=3c^2 M_p^{2}L^{-2}$ where $M_p = 1/\sqrt{8\pi G}$
is the reduced Planck mass and $c$ is a numerical constant
characterizing all of the uncertainties of the theory, whose value
can only be determined by observations. On the basis of the
holographic principle proposed by [13] several others have studied
holographic model for dark energy [14]. Employment of Friedman
equation $\rho=3M_{p}^{2}H^{2}$ where $\rho$ is the total energy
density and taking $L=H^{-1}$ one can find
$\rho_{m}=3(1-c^{2})M_{p}^{2}H^{2}$. Thus either $\rho_{m}$ or
$\rho_{\Lambda}$ behaves like $H^{2}$. If we take $L$ as the size
of the current universe, say, the Hubble radius $\frac{1}{H}$ then
the dark energy density will be close to the observational result.
There are other type of dark energy i.e., Ricci dark energy, which
is a kind of holographic dark energy [15] taking the square root
of the inverse Ricci scalar as its infrared cutoff and this model
is also phenomenologically viable. Gao et al [16] proposed the
dark energy density proportional to the Ricci scalar $R$ i.e.,
$\rho_{X}\propto R$. This model works fairly well in fitting the
observational data, and it could also help to understand the
coincidence problem. There are several works on this
Ricci dark energy model [17]. \\

Here we shall consider two types of dark energy models: (i) generalized
holographic and (ii) generalized Ricci dark energy models [18]. In Section II,
we have found some solutions if the universe is filled with dark matter and
generalized holographic/Ricci dark energy. General prescriptions of first law and
GSL of thermodynamics have been presented in section III. We have shown
that the first law and GSL are always valid on apparent horizon
and first law can not be satisfied on the particle and event horizons in Einstein's gravity. These
results are always true for any types of dark energy models i.e., these results do
not depend on the dark energy models in Einstein's gravity. But the GSL completely depends
on the choices of dark energy models in Einstein's gravity.
There are several works on the validity of GSL on the particle and event horizons in various
dark energy models [19, 20, 21]. Here we shall discuss the validity of GSL in
Generalized holographic and generalized Ricci dark energy models in section IV. Also
if we consider the generalized holographic dark energy and generalized Ricci dark
energy as the original holographic dark energy, then our aim is to check the validity
of laws of thermodynamics on the horizon of radius $L$. Some conclusions have been
drawn in last section.\\

\section{\normalsize\bf{Generalized Holographic and Ricci Dark Energy Models}}

The Einstein's field equations for homogeneous, isotropic and flat
FRW universe are given by

\begin{equation}
H^{2}=\frac{8\pi}{3}\rho
\end{equation}
and
\begin{equation}
\dot{H}=-4\pi(\rho+ p)
\end{equation}

where $H(=\frac{\dot{a}}{a})$ is the Hubble parameter (choosing
$G=c=1$) and $\rho$ and $p$ be the energy density and pressure of
the fluid respectively. The conservation equation is given by

\begin{equation}
\dot{\rho}+3H(\rho+p)=0
\end{equation}

Now let us consider the fluid is a combination of dark matter and
dark energy i.e., $\rho=\rho_{m}+\rho_{DE}$ and $p=p_{m}+p_{DE}$
with $p_{m}=0$. Assuming there is no interaction between dark
matter and dark energy, so they are separately conserved. Thus the
the conservation equation (3) becomes

\begin{equation}
\dot{\rho}_{m} + 3H\rho_{m}=0
\end{equation}

and

\begin{equation}
\dot{\rho}_{DE} + 3H(\rho_{DE}+ p_{DE})=0
\end{equation}

From equation (4), we get the energy density of matter as

\begin{equation}
\rho_{m}=\rho_{m_{0}} (1+z)^{ 3}
\end{equation}

where $\rho_{m_{0}}$ is an integration constant which gives the
present value of the dark energy density and $z=\frac{1}{a}-1$ is the redshift.\\

Recently, Xu et al [18] proposed two types of dark energy models, i.e.,
generalized holographic and generalized Ricci dark energy models
as follows:\\

(i) {\bf{Generalized Holographic Dark Energy Model (GHDEM):}} The
energy density of GHDEM is given by,

\begin{equation}
\rho_{h}=\rho_{DE}=\frac{3c^2}{8\pi}H^2f(R/H^2)
\end{equation}

where $c$ is a numerical constant and $f(x)$ is a positive
function defined as, $f(x)=\alpha x+(1-\alpha)$, $\alpha$ is a
constant. Here $R$ is the Ricci scalar and its expression for flat
universe is given by

\begin{equation}
R=-6(\dot{H}+2H^2)
\end{equation}

When $\alpha=0$ then $f(x)=1$, we recover the energy densities of
original holographic dark energy. Also when $\alpha=1$
then $f(x)=x$, we recover the energy density of
original Ricci dark energy.\\

So the energy density $\rho_{h}$ becomes

\begin{equation}
\rho_{h}=\frac{3c^2}{8 \pi}[(1-13\alpha) H^{2} -6\alpha\dot{H}]
\end{equation}

Solving $(1)$, $(6)$ and $(9)$ we get

\begin{equation}
H^{2}=\frac{8\pi\rho_{m_{0}}(1+z)^{ 3}}{3(1 - c^{2} + 4 \alpha
c^{2})} +H_{0}^{2}~(1+z)^{\frac{c^{2}(13\alpha-1)+1}{3\alpha c^{2}}}
\end{equation}

where, $H_{0}$ is an integration constant. Differentiating (10) w.r.t. $t$, we get

\begin{equation}
\dot{H}=-\frac{8\pi\rho_{m_{0}}(1+z)^{ 3}}{2(1 - c^{2} + 4 \alpha
c^{2})} + \frac{ c^{2}(1 - 13 \alpha )- 1}{6 \alpha c^{2}}~H_{0}^{2}~ (1+z)^{\frac{c^{2}(13\alpha-1)+1}{3\alpha c^{2}}}
\end{equation}

 From equation $(8)$, we obtain the Ricci scalar $R$ as,

\begin{equation}
R=-\frac{8\pi\rho_{m_{0}}(1+z)^{ 3}}{(1 - c^{2} + 4 \alpha c^{2})}
- \frac{ c^{2}(1 - 7 \alpha )- 1}{3 \alpha c^{2}} ~H_{0}^{2}~(1+z)^{\frac{c^{2}(13\alpha-1)+1}{3\alpha c^{2}}}
\end{equation}

So from equation (1) and (2), we can found the expressions for
density $\rho_{h}$ and pressure $p_{h}$ as,

\begin{equation}
\rho_{h}=\frac{c^{2}(1-4\alpha)\rho_{m_{0}}(1+z)^{ 3}}{(1 - c^{2}
+ 4 \alpha c^{2})} +\frac{3}{8\pi}~H_{0}^{2}~(1+z)^{\frac{c^{2}(13\alpha-1)+1}{3\alpha c^{2}}}
\end{equation}

and

\begin{equation}
p_{h}= \frac{ c^{2}(4 \alpha -1)+ 1}{24\pi
\alpha c^{2}} ~H_{0}^{2}~(1+z)^{\frac{c^{2}(13\alpha-1)+1}{3\alpha c^{2}}}
\end{equation}\\

(ii) {\bf{Generalized Ricci Dark Energy Model (GRDEM):}} The
energy density of GRDEM is given by,

\begin{equation}
\rho_{r}=\frac{3c^2}{8\pi}R~g(H^2/R)
\end{equation}

where  $g(y)$ is a positive function defined as, $g(y)=\beta
y+(1-\beta)$, $\beta$ is a constant. When $\beta=0$ then $g(y)=1$,
we recover the energy density of
original Ricci dark energy. Also when $\beta=1$
then $g(y)=x$, we recover the energy density of
original holographic dark energy.\\

Now comparing (7) and (15), we see that when $\beta=1-\alpha$ the generalized Ricci
dark energy reduces to the generalized holographic dark energy and vice versa. If we
replace $\alpha$ by $(1-\beta)$ in equations (9) - (14), we will get similar solutions for
generalized Ricci dark energy model. Therefore, equations (9) - (14) are also the solutions
of generalized Ricci dark energy model with $\alpha=1-\beta$.\\

\section{\normalsize\bf{First Law and Generalized Second Law of Thermodynamics (GSL): General Prescription}}

We consider the FRW universe as a thermodynamical system with the
horizon surface as a boundary of the system. Here, we study the
validity of first law and generalized second law (GSL) of thermodynamics
on apparent, particle and event horizons. To study the
generalized second law (GSL) of thermodynamics through the
universe we deduce the expression for normal entropy using the
Gibb's law of thermodynamics [22]

\begin{equation}
T_{X}dS_{I}=pdV+d(E_{X})
\end{equation}

where, $S_{I},~p,~V$ and $E_{X}$ are respectively entropy,
pressure, volume and internal energy within the apparent/particle/event
horizon and $T_{X}$ is the temperature on the apparent horizon
($X=A$)/particle horizon ($X=P$)/event horizon ($X=E$). Here the expression for internal
energy can be written as $E_{X}=\rho V$. Now the volume of the
sphere is $V=\frac{4}{3}\pi R_{X}^{3}$, where $R_{X}$ is the
radius of the apparent horizon ($R_{A}$)/particle horizon ($R_{P}$)/event horizon ($R_{E}$)
defined by [22] (see also our previous work [24])

\begin{equation}
R_{A}=\frac{1}{H}~,
\end{equation}

\begin{equation}
R_{P}=a\int_{0}^{t}\frac{dt}{a}=\frac{1}{1+z}\int_{z}^{\infty}\frac{dz}{H}
\end{equation}

and
\begin{equation}
R_{E}=a\int_{t}^{\infty}\frac{dt}{a}=\frac{1}{1+z}\int_{-1}^{z}\frac{dz}{H}
\end{equation}

which immediately give (after differentiation)

\begin{equation}
\dot{R}_{P}=HR_{P}+1
\end{equation}
and
\begin{equation}
\dot{R}_{E}=HR_{E}-1
\end{equation}

The temperature and the entropy on the apparent/particle/event horizon are also given by

\begin{equation}
T_{X}=\frac{1}{2\pi R_{X}}
\end{equation}
and
\begin{equation}
S_{X}=\pi R_{X}^{2}
\end{equation}

Next, we shall examine the validity of first law and GSL on the apparent horizon, particle horizon and event horizon.\\\\

$\bullet$ {\bf First law on the apparent horizon, particle horizon and event horizon:}\\

The amount of the energy crossing on the apparent/particle/event horizon is
[23] given by

\begin{equation}
-dE_{X}=4\pi R_{X}^{3}HT_{\mu\nu}k^{\mu}k^{\nu}dt=4\pi
R_{X}^{3}H(\rho+p)dt=-H\dot{H}R_{X}^{3}dt
\end{equation}

The first law of thermodynamics on the apparent/particle/event horizon is
defined as follows:

\begin{equation}
-dE_{X}=T_{X}dS_{X}
\end{equation}

On the apparent horizon, we have (using (17), (22), (23) and (24))

\begin{equation}
-dE_{A}=-H\dot{H}R_{A}^{3}dt=-\frac{\dot{H}}{H^{2}}dt
\end{equation}

and

\begin{equation}
T_{A}dS_{A}=-\dot{R}_{A}dt=-\frac{\dot{H}}{H^{2}}dt
\end{equation}

The equations (26) and (27) imply,

\begin{equation}
-dE_{A}=T_{A}dS_{A}
\end{equation}

This shows that the first law is always satisfied on the apparent horizon in Einstein's gravity.\\

On the particle horizon, we have (using (18), (20), (22), (23) and (24))

\begin{equation}
-dE_{P}=-H\dot{H}R_{P}^{3}dt
\end{equation}

and

\begin{equation}
T_{P}dS_{P}=\dot{R}_{P}dt=(HR_{P}+1)dt
\end{equation}

From equations (29) and (30), we get

\begin{equation}
-dE_{P}=T_{P}dS_{P}-(H\dot{H}R_{P}^{3}+HR_{P}+1)dt
\end{equation}

The second term on the r.h.s. of (31) is time dependent, so this term
cannot become zero during certain stage of the evolution of the universe. Thus
we may conclude that

\begin{equation}
-dE_{P}\ne T_{P}dS_{P}
\end{equation}

So on the particle horizon, first law of thermodynamics can not be
satisfied in Einstein's gravity.\\

On the event horizon, we have (using (19), (21), (22), (23) and (24))

\begin{equation}
-dE_{E}=-H\dot{H}R_{E}^{3}dt
\end{equation}

and

\begin{equation}
T_{E}dS_{E}=\dot{R}_{E}dt=(HR_{E}-1)dt
\end{equation}

From equations (33) and (34), we get

\begin{equation}
-dE_{E}=T_{E}dS_{E}-(H\dot{H}R_{E}^{3}+HR_{E}-1)dt
\end{equation}

The second term on the r.h.s. is time dependent, so this term
cannot become zero during certain stage of the evolution of the universe. Thus
we may conclude that

\begin{equation}
-dE_{E}\ne T_{E}dS_{E}
\end{equation}

So on the event horizon, first law of thermodynamics can not be
satisfied in Einstein's gravity.\\\\\\

$\bullet$ {\bf GSL on the apparent horizon, particle horizon and event horizon:}\\

The rate of change of internal entropy and total entropy are obtained
as (from (16), (22) and (23))

\begin{equation}
\dot{S}_{I}=\frac{\dot{H}R_{X}^{2}(HR_{X}-\dot{R}_{X})}{T_{X}}
\end{equation}
and
\begin{equation}
\dot{S}_{I}+\dot{S}_{X}=2\pi R_{X}[\dot{H}R_{X}^{2}(HR_{X}-\dot{R}_{X})+\dot{R}_{X}]
\end{equation}

The GSL states that total entropy can not be
decreased i.e.,

\begin{equation}
\dot{S}_{I}+\dot{S}_{X}\ge 0 ~i.e.,~\dot{H}R_{X}^{2}(HR_{X}-\dot{R}_{X})+\dot{R}_{X}
\ge 0
\end{equation}

On the apparent horizon, the rate of change of total entropy is (using (17) and (38))

\begin{equation}
\dot{S}_{I}+\dot{S}_{A}=2\pi R_{A}[\dot{H}R_{A}^{2}(HR_{A}-\dot{R}_{A})+\dot{R}_{A}]=\frac{2\pi\dot{H}^{2}}{H^{5}}
\ge 0
\end{equation}

So GSL always satisfied on the apparent horizon in Einstein's gravity.\\

On the particle horizon, the rate of change of total entropy is (using (20) and (38))

\begin{equation}
\dot{S}_{I}+\dot{S}_{P}=2\pi R_{P}[\dot{H}R_{P}^{2}(HR_{P}-\dot{R}_{P})+\dot{R}_{P}]=2\pi
R_{P}\left(-\dot{H}R_{P}^{2}+HR_{P}+1 \right)
\end{equation}

So GSL may be satisfied on the particle horizon if the following
condition holds:

\begin{equation}
-\dot{H}R_{P}^{2}+HR_{P}+1 \ge 0
\end{equation}

For quintessence model, $\dot{H}<0$, so GSL always satisfied. But for phantom model, $\dot{H}>0$,
GSL may or may not be satisfied. For phantom crossing model, the GSL satisfied in the initial stage but
in the late stage the GSL may not be satisfied during evolution of the universe on the particle horizon.\\

On the event horizon, the rate of change of total entropy is (using (21) and (38))

\begin{equation}
\dot{S}_{I}+\dot{S}_{E}=2\pi R_{E}[\dot{H}R_{E}^{2}(HR_{E}-\dot{R}_{E})+\dot{R}_{E}]=2\pi
R_{E}\left(\dot{H}R_{E}^{2}+HR_{E}-1 \right)
\end{equation}

So GSL may be satisfied on the event horizon if the following
condition holds:

\begin{equation}
\dot{H}R_{E}^{2}+HR_{E}-1 \ge 0
\end{equation}

For quintessence model ($\dot{H}<0$) or phantom model $\dot{H}>0$ or
phantom crossing model, we can not draw any definite conclusion for validity of GSL
on event horizon. \\

So from above discussions, we may conclude that the validity of first
law is independent of dark energy model in Einstein's gravity on apparent horizon, particle horizon and
event horizon and also validity of GSL is independent of dark energy model in
Einstein's gravity on apparent horizon. But validity of GSL is
completely depends on the dark energy models in Einstein's gravity on particle and event horizons. Hence in the
next section we shall discuss the validity of GSL on particle and event horizons for generalized
holographic and generalized Ricci dark energy models.\\

\begin{figure}
\includegraphics[height=2.2in]{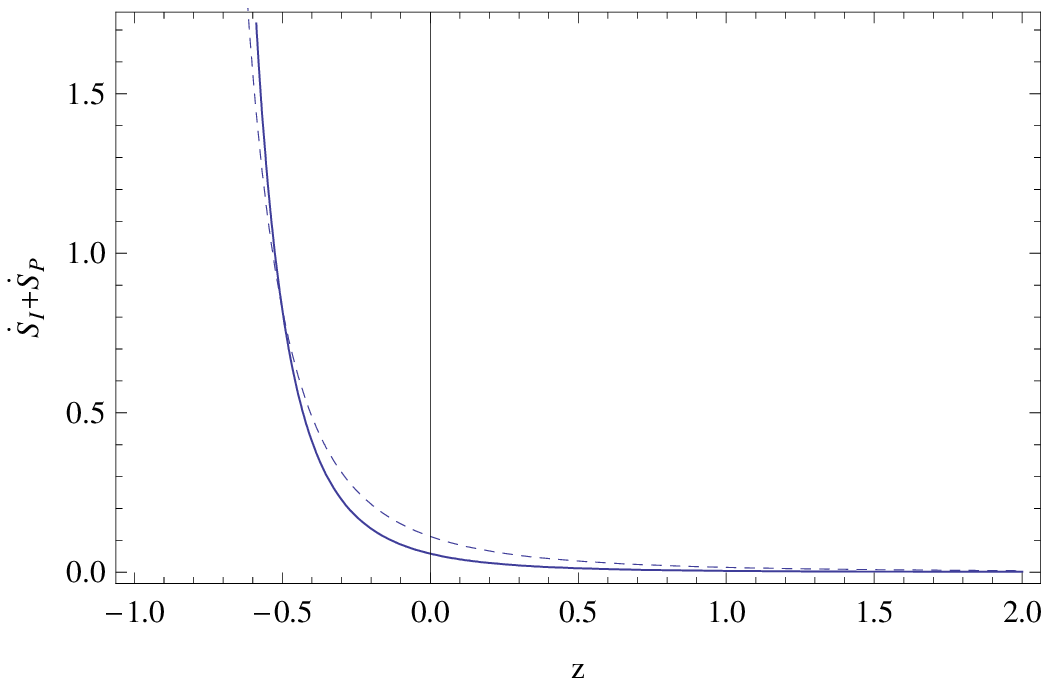}~~~~
\includegraphics[height=2.2in]{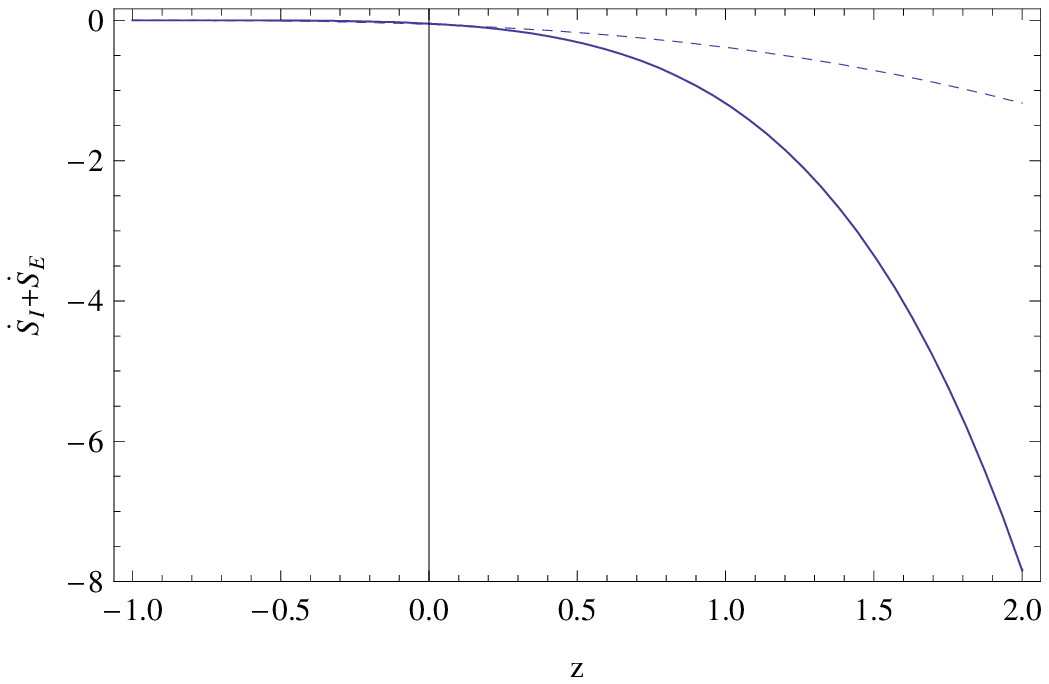}\\
\vspace{1mm} ~~~~~Fig.1~~~~~~~~~~~~~~~~~~~~~~~~~~~~~~~~~~~~~~~~~~~~~~~~~~~~~~~Fig.2\\

\vspace{6mm} Figs. 1 and 2 show the variations of $(\dot{S}_{I}+\dot{S}_{P})$
and $(\dot{S}_{I}+\dot{S}_{E})$ respectively against
redshift $z$ for $c=.5,\rho_{m_{0}}=1,\alpha=.7,\beta=.7,H_{0}=70$. The dash line and
non-dash line represent for the generalized holographic and generalized Ricci dark energy models
respectively.

 \vspace{10mm}

\end{figure}

\section{\normalsize\bf{Validity of GSL on Particle and Event Horizons in Generalized Holographic and Ricci Dark Energy Models}}

In generalized holographic/Ricci dark energy model, the radii of the particle and event horizons in terms of
redshift $z$ can be written as (using (10), (18) and (19))

\begin{equation}
R_{P}=\frac{1}{1+z}\int_{z}^{\infty}\frac{dz}{H}=\frac{1}{1+z}\int_{z}^{\infty} \left[\frac{8\pi\rho_{m_{0}}(1+z)^{ 3}}{3(1 - c^{2} + 4 \alpha
c^{2})} +H_{0}^{2}~(1+z)^{\frac{c^{2}(13\alpha-1)+1}{3\alpha c^{2}}} \right]^{-\frac{1}{2}} ~dz
\end{equation}

and

\begin{equation}
R_{E}=\frac{1}{1+z}\int_{-1}^{z}\frac{dz}{H}=\frac{1}{1+z}\int_{-1}^{z} \left[\frac{8\pi\rho_{m_{0}}(1+z)^{ 3}}{3(1 - c^{2} + 4 \alpha
c^{2})} +H_{0}^{2}~(1+z)^{\frac{c^{2}(13\alpha-1)+1}{3\alpha c^{2}}} \right]^{-\frac{1}{2}} ~dz
\end{equation}\\

On the particle horizon, the rate of change of total entropy is (using (10) and (41)) obtained as

\begin{eqnarray*}
\dot{S}_{I}+\dot{S}_{P}=2\pi
R_{P}\left(-\dot{H}R_{P}^{2}+HR_{P}+1 \right)~~~~~~~~~~~~~~~~~~~~~~~~~~~~~~~~~~~~~~~~~~~~~~~~~~~~~~~~~~~~~~~~~~~~~~~~~~~~
\end{eqnarray*}

\begin{eqnarray*}
=2\pi
R_{P}\left[\left\{\frac{8\pi\rho_{m_{0}}(1+z)^{ 3}}{2(1 - c^{2} + 4 \alpha
c^{2})} - \frac{ c^{2}(1 - 13 \alpha )- 1}{6 \alpha c^{2}}
~H_{0}^{2}~ (1+z)^{\frac{c^{2}(13\alpha-1)+1}{3\alpha c^{2}}}\right\}  R_{P}^{2}\right.
\end{eqnarray*}

\begin{equation}
\left.+\left\{\frac{8\pi\rho_{m_{0}}(1+z)^{ 3}}{3(1 - c^{2} + 4 \alpha
c^{2})} +H_{0}^{2}~(1+z)^{\frac{c^{2}(13\alpha-1)+1}{3\alpha c^{2}}} \right\}^{\frac{1}{2}}R_{P}~~+1 \right]
\end{equation}\\

Also, on the event horizon, the rate of change of total entropy is (using (10) and (43)) obtained as

\begin{eqnarray*}
\dot{S}_{I}+\dot{S}_{E}=2\pi
R_{E}\left(\dot{H}R_{E}^{2}+HR_{E}-1 \right)~~~~~~~~~~~~~~~~~~~~~~~~~~~~~~~~~~~~~~~~~~~~~~~~~~~~~~~~~~~~~~~~~~~~~~~~~~~~
\end{eqnarray*}

\begin{eqnarray*}
=2\pi
R_{E}\left[\left\{-\frac{8\pi\rho_{m_{0}}(1+z)^{ 3}}{2(1 - c^{2} + 4 \alpha
c^{2})} + \frac{ c^{2}(1 - 13 \alpha )- 1}{6 \alpha c^{2}}
~H_{0}^{2}~ (1+z)^{\frac{c^{2}(13\alpha-1)+1}{3\alpha c^{2}}}\right\}  R_{E}^{2}\right.
\end{eqnarray*}

\begin{equation}
\left.+\left\{\frac{8\pi\rho_{m_{0}}(1+z)^{ 3}}{3(1 - c^{2} + 4 \alpha
c^{2})} +H_{0}^{2}~(1+z)^{\frac{c^{2}(13\alpha-1)+1}{3\alpha c^{2}}} \right\}^{\frac{1}{2}}R_{E}~~-1 \right]
\end{equation}\\

The rate of change of total entropies for particle and event horizons have
drawn in figs. 1 and 2 respectively. So Figs. 1 and 2 show the variations of $(\dot{S}_{I}+\dot{S}_{P})$
and $(\dot{S}_{I}+\dot{S}_{E})$ respectively against
redshift $z$ for $c=.5,\rho_{m_{0}}=1,\alpha=.7,\beta=.7,H_{0}=70$. The dash line and
non-dash line represent for the generalized holographic and generalized Ricci dark energy models
respectively. From the figures, we see that the GSL may be satisfied on the particle horizon for
generalized holographic and Ricci dark energy models, but on the event horizon GSL can not be satisfied
for these two dark energy models.\\

\section{\normalsize\bf{Generalized Holographic/Ricci as an Original Holographic Dark Energy Model: Thermodynamics}}

We consider here that the generalized holographic/Ricci dark energy model
as an original holographic dark energy model. For this purpose, we will
compare the energy densities between generalized holographic/Ricci and
original holographic dark energies. Now if we compare the energy density
of holographic dark energy which is given in (7) with the original holographic
model with energy density,

\begin{equation}
\rho_{\Lambda}=\frac{3c^2}{8 \pi}L^{-2}
\end{equation}

then we get,

\begin{equation}
L^{2}=\frac{1}{\alpha R +(1 - \alpha)H^{2}}
\end{equation}

which can be written in the simplified form (using (10) and (12)):

\begin{equation}
L^{2}=\frac{3 c^{2} (1 - c^{2} + 4 \alpha c^{2})}{8 c^{2}(1 -
4\alpha) \pi \rho_{m_{0}}(1+z)^{ 3} + [2c^{2}(1 + 2 \alpha)+ 1]
(1 - c^{2} + 4 \alpha c^{2}) H_{0}^{2}~(1+z)^{\frac{c^{2}(13\alpha-1)+1}{3\alpha c^{2}}}}
\end{equation}\\

Here, we will try to apply the usual definition of the
temperature and entropy on the horizon of radius $L$ and
examine the validity of first and the second laws of thermodynamics.
The temperature and the entropy on the horizon are (similar to (22) and (23)) [25]

\begin{equation}
T_{L}=\frac{1}{2\pi L}
\end{equation}
and
\begin{equation}
S_{L}=\pi L^{2}
\end{equation}

The amount of the energy crossing on the horizon is (same as (24)) given by [25]

\begin{equation}
-dE_{L}=4\pi L^{3}HT_{\mu\nu}k^{\mu}k^{\nu}dt=4\pi
L^{3}H(\rho+p)dt=-H\dot{H}L^{3}dt
\end{equation}

and from (52) and (53) we have

\begin{equation}
T_{L}dS_{L}=\dot{L}dt
\end{equation}

From (54) and (55), we obtain the relation

\begin{equation}
-dE_{L}=T_{L}dS_{L}-(H\dot{H}L^{3}+\dot{L})dt
\end{equation}

The second term on the r.h.s. is time dependent, so this term
cannot become zero during certain stage of the evolution of the universe. Thus
we may conclude that

\begin{equation}
-dE_{L}\ne T_{L}dS_{L}
\end{equation}

So on the horizon, first law of thermodynamics can not be
satisfied in Einstein's gravity.\\

\begin{figure}
\includegraphics[height=2.5in]{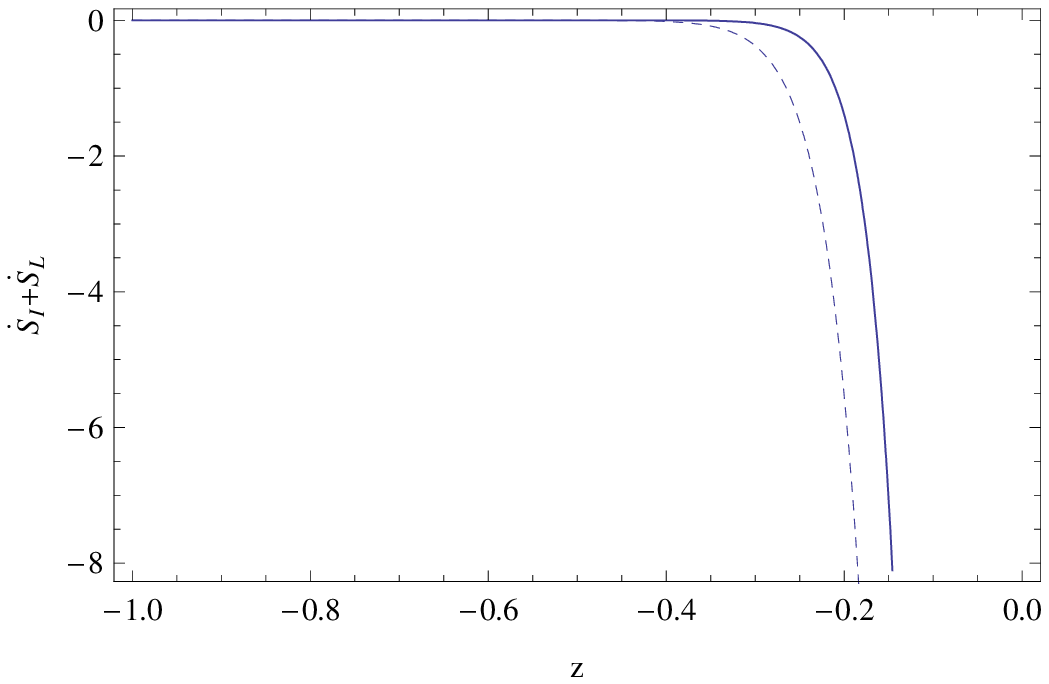}\\
\vspace{1mm} ~~~~~Fig.3~~~~~~~~~\\

\vspace{6mm} Fig. 3 shows the variation of $(\dot{S}_{I}+\dot{S}_{L})$ against
redshift $z$ for $c=.5,\rho_{m_{0}}=1,\alpha=.7,\beta=.7,H_{0}=70$. The dash line and
non-dash line represent for the generalized holographic and generalized Ricci dark energy models
respectively.

 \vspace{10mm}

\end{figure}

Now on the horizon, the rate of change of total entropy is (similar to (38)) obtained as

\begin{equation}
\dot{S}_{I}+\dot{S}_{L}=2\pi L[\dot{H}L^{2}(HL-\dot{L})+\dot{L}]
\end{equation}

The GSL will be satisfied if

\begin{equation}
\dot{H}L^{2}(HL-\dot{L})+\dot{L}\ge 0
\end{equation}

For quintessence model ($\dot{H}<0$) or phantom model $\dot{H}>0$ or
phantom crossing model, we can not draw any definite conclusion for validity of GSL
on the horizon. The validity is completely depends on the value of $L$ in the
dark energy model. In generalized holographic dark energy model,
the rate of change of total entropy on the horizon is calculated as
(from (10), (11) and (58))

\begin{eqnarray*}
\dot{S}_{I}+\dot{S}_{L}=\pi H \left[\dot{H}L^{2}\left(2L^{2}+(1+z)\frac{dL^{2}}{dz}\right)
-(1+z)\frac{dL^{2}}{dz}\right]~~~~~~~~~~~~~~~~~~~~~~~~~~~~~~~~~~~~
\end{eqnarray*}

\begin{eqnarray*}
=\pi \left[\frac{8\pi\rho_{m_{0}}(1+z)^{ 3}}{3(1 - c^{2} + 4 \alpha
c^{2})} +H_{0}^{2}~(1+z)^{\frac{c^{2}(13\alpha-1)+1}{3\alpha c^{2}}}\right]^{\frac{1}{2}} \times
\left[-(1+z)\frac{dL^{2}}{dz} +\right.
\end{eqnarray*}

\begin{equation}
\left. +\left\{
-\frac{8\pi\rho_{m_{0}}(1+z)^{ 3}}{2(1 - c^{2} + 4 \alpha
c^{2})} + \frac{ c^{2}(1 - 13 \alpha )- 1}{6 \alpha c^{2}}
~H_{0}^{2}~ (1+z)^{\frac{c^{2}(13\alpha-1)+1}{3\alpha c^{2}}}\right\}
L^{2}\left(2L^{2}+(1+z)\frac{dL^{2}}{dz}\right)\right]
\end{equation}\\

where $L^{2}$ has the expression given in equation (51). This is very complicated
expression in $z$, so we need graphical investigation. The rate of change of total
entropy on the horizon of radius $L$ have
drawn in fig. 3. So Fig. 3 shows the variation of $(\dot{S}_{I}+\dot{S}_{L})$
against redshift $z$ for $c=.5,\rho_{m_{0}}=1,\alpha=.7,\beta=.7,H_{0}=70$. The dash line and
non-dash line represent for the generalized holographic and generalized Ricci dark energy models
respectively. From the figure, we see that the GSL can not be satisfied on the horizon for
generalized holographic and Ricci dark energy models.\\

\section{\normalsize\bf{Discussions}}

We have considered the flat FRW model of the universe which is
filled with the combination of dark matter and dark energy. Here
we have considered two types of dark energy models: (i) Generalized
holographic and (ii) generalized Ricci dark energies. If $\beta=1-\alpha$,
the generalized Ricci dark energy model is converted to generalized
holographic dark energy model. When $\alpha=0$ or $\beta=1$, we recover
the energy densities of original holographic dark energy. Also when
$\alpha=1$ or $\beta=0$, we recover the energy density of original
Ricci dark energy. In these two models, the solutions have been found
in terms of the redshift $z$. The general descriptions of first law and
generalized second law (GSL) of thermodynamics have studied on the
apparent horizon, particle horizon and event horizon of the universe.
Here we have tried to apply the usual definition of the
temperature and entropy as that of the apparent horizon to the particle
horizon and the cosmological event horizon and examine the validity of first and
the second laws of thermodynamics. We have shown that the first law and GSL
are always valid on apparent horizon and first law can not be satisfied
on the particle and event horizons in Einstein's gravity.
These results are always true for any types of dark energy models i.e.,
these results do not depend on the dark energy models in Einstein's gravity.
But the GSL completely depends on the choices of dark energy models in Einstein's
gravity. Here we have discussed the validity of GSL in Generalized holographic
and generalized Ricci dark energy models. From figures 1 and 2, we have seen that
the GSL may be satisfied on the particle horizon and can not be satisfied on the
event horizon for both generalized holographic and Ricci dark energy models.
Also we have considered the Generalized holographic dark energy and generalized
Ricci dark energy as the original holographic dark energy, so in this situation
we have calculated the expression of the radius of the horizon $L$ in terms of
redshift $z$. On this horizon, we have shown that the first law can not be satisfied.
From figure 3, we have seen that the GSL can not be satisfied on the horizon of radius
$L$ for both dark energy models.\\

{\bf Acknowledgement:}\\

The authors are thankful to IUCAA, Pune, India for warm
hospitality where part of the work was carried out.\\

{\bf References:}\\
\\
$[1]$ S. J. Perlmutter et al, {\it Bull. Am. Astron. Soc.} {\bf
29} 1351 (1997); S. Perlmutter et al, {\it Nature} {\bf 391} 51
(1998); S. J. Perlmutter et al, \textit{Astrophys. J.} \textbf{517} 565 (1999).\\
$[2]$ A. G. Riess et al, {\it Astron. J.} {\bf 116} 1009 (1998);
P. Garnavich et al, {\it Astrophys. J.} {\bf 493} L53 (1998);
B. P. Schmidt et al, {\it Astrophys. J.} {\bf 507} 46 (1998);
N. A. Bachall, J. P. Ostriker, S. Perlmutter and P. J. Steinhardt,
\textit{Science} \textbf{284} 1481 (1999).\\
$[3]$ V. Sahni and A. A. Starobinsky, {\it Int. J. Mod. Phys. A}
{\bf 9} 373 (2000); P. J. E. Peebles and B. Ratra, {\it Rev. Mod. Phys.} {\bf
75} 559 (2003).\\
$[4]$ T. Padmanabhan, {\it Phys. Rept.} {\bf 380} 235 (2003);
E. J. Copeland, M. Sami, S. Tsujikawa, {\it Int. J. Mod.
Phys. D} {\bf  15} 1753 (2006); J. A. Frieman, M. S. Turner and
D. Huterer, arXiv:0803.0982[astro-ph].\\
$[5]$ B. Ratra and P. J. E. Peebles, {\it Phys. Rev. D} {\bf
37} 3406 (1988).\\
$[6]$ T. Chiba, T. Okabe and M. Yamaguchi, {\it Phys. Rev. D}
{\bf 62} 023511 (2000).\\
$[7]$ A. Sen, {\it JHEP} {\bf 0204} 048 (2002).\\
$[8]$ A. Kamenshchik, U. Moschella and V. Pasquier, {\it Phys.
Lett. B} {\bf 511} 265 (2001); V. Gorini, A. Kamenshchik, U.
Moschella and V. Pasquier, {\it gr-qc}/0403062; V. Gorini, A.
Kamenshchik and U. Moschella, {\it Phys. Rev. D} {\bf 67} 063509
(2003); U. Alam, V. Sahni, T. D. Saini and A. A. Starobinsky, {\it
Mon. Not. R. Astron. Soc.} {\bf 344}, 1057 (2003); H. B. Benaoum,
{\it hep-th}/0205140; U. Debnath, A. Banerjee and S. Chakraborty,
{\it Class.
Quantum Grav.} {\bf 21} 5609 (2004).\\
$[9]$ R. R. Caldwell, {\it Phys. Lett. B} {\bf 545} 23 (2002).\\
$[10]$ E. Witten, {\it hep-ph}/0002297.\\
$[11]$ A. G. Cohen, D. B. Kaplan and A. E. Nelson, {\it Phys. Rev.
Lett.} {\bf 82} 4971 (1999).\\
$[12]$ X. Zhang, {\it Int. J. Mod. Phys. D} {\bf 14} 1597 (2005).\\
$[13]$ W. Fischler and L. Susskind, {\it hep-th}/9806039.\\
$[14]$ M. R. Setare, {\it Phys. Lett. B} {\bf 648} 329 (2007);
Y. Gong, {\it Phys. Rev. D} {\bf 70} 064029 (2004).\\
$[15]$ M. Li, {\it Phys. Lett. B} {\bf 603} 1 (2004).\\
$[16]$ C. Gao, F. Wu, X. Chen and Y. -G. Shen, {\it Phys. Rev. D}
{\bf 79} 043511 (2009).\\
$[17]$ C. -J. Feng, {\it Phys. Lett. B} {\bf 670} 231 (2008); {\it Phys. Lett. B}
{\bf 672} 94 (2009); arXiv: 0806:0673 [hep-th]; C. -J. Feng and X. -Z. Li, {\it Phys. Lett. B} {\bf 680} 355 (2009);
L. Xu, W. Li and J. Lu, {\it Mod. Phys. Lett. A} {\bf 24} 1355 (2009); M. Suwa and T. Nihei,
{\it Phys. Rev. D} {\bf 81} 023519 (2010); K. W. Kim, H. W. Lee and Y. S. Myung, arXiv: 0812:4098 [gr-qc].\\
$[18]$ L. Xu, J. Lu and W. Li., {\it Eur. Phys. J. C} {\bf 64} 89 (2009).\\
$[19]$ H. M. Sadjadi, {\it JCAP} {\bf 02} 026 (2007); H. M. Sadjadi and M. Honardoost,
{\it Phys. Lett. B} {\bf 647} 231 (2007).\\
$[20]$ K. Karami and A. Abdolmaleki, arXiv: 0909.2427 [gr-qc]; K. Karami and S. Ghaffari, {\it Phys. Lett. B} {\bf 685} 115 (2010).\\
$[21]$ Y. Zhang, Z. Yi, T. -J. Zhang and W. Liu, {\it Phys. Rev. D} {77} 023502 (2008).\\
$[22]$ B. Wang, Y. G. Gong and E. Abdalla, {\it Phys. Rev. D} {\bf
74} 083520 (2006); G. Izquierdo and D. Pavon, {\it Phys. Lett. B}
{\bf 633} 420 (2006).\\
$[23]$ R. Bousso, {\it Phys. Rev. D} {\bf 71} 064024 (2005).\\
$[24]$ S. Bhattacharya and U. Debnath, arXiv: 1006.2600[gr-qc]; arXiv: 1006.2609[gr-qc].\\
$[25]$ M. R. Setare and S Shafei, {\it JCAP} {\bf 09} 011 (2006);
M. R. Setare, {\it JCAP} {\bf 01} 023 (2007).\\

\end{document}